\documentclass[aps,prl,twocolumn,amsmath,amssymb,nofootinbib,superscriptaddress,floatfix,reprint,longbibliography]{revtex4-1}

\usepackage{graphicx}
\usepackage{dcolumn}
\usepackage{bm}
\usepackage{tabularx}
\usepackage{amsmath}
\usepackage{textcomp}
\usepackage{upgreek}
\usepackage{color}
\makeatletter

\newcommand{\Rmnum}[1]{\expandafter\@slowromancap\romannumeral #1@}
\makeatother

\usepackage[colorlinks=true, urlcolor=blue, linkcolor=blue, citecolor=blue, pdftex]{hyperref}
\usepackage{ulem} 
\usepackage[pdftex]{color} 

\begin{document}

\title{Fermi surface nesting, vacancy ordering and the emergence of superconductivity in IrSb compounds}
\author{Tianping Ying$^\dagger$}\affiliation{Materials Research Center for Element Strategy, Tokyo Institute of Technology, Yokohama 226-8503, Japan} \affiliation{Beijing National Laboratory for Condensed Matter Physics, Institute of Physics, Chinese Academy of Sciences, Beijing 100190, China}
\author{,Tongxu Yu$^\dagger$}\affiliation{Gusu Laboratory of Materials, Jiangsu 215123, China}
\author{Erjian Cheng}\affiliation{ Department of Physics, Fudan University, Shanghai 200438, China}
\author{Shiyan Li}\affiliation{	Department of Physics, Fudan University, Shanghai 200438, China}
\author{Jun Deng}\affiliation{Beijing National Laboratory for Condensed Matter Physics, Institute of Physics, Chinese Academy of Sciences, Beijing 100190, China}
\author{Xiangru Cui}\affiliation{Beijing National Laboratory for Condensed Matter Physics, Institute of Physics, Chinese Academy of Sciences, Beijing 100190, China}
\author{Jiangang Guo}\affiliation{Beijing National Laboratory for Condensed Matter Physics, Institute of Physics, Chinese Academy of Sciences, Beijing 100190, China}
\author{Yanpeng Qi$^*$}\affiliation{School of Physical Science and Technology, ShanghaiTech University, Shanghai 201210, China} 
\author{Xiaolong Chen$^*$}\affiliation{Beijing National Laboratory for Condensed Matter Physics, Institute of Physics, Chinese Academy of Sciences, Beijing 100190, China}
\author{Hideo Hosono$^{*,}$}\affiliation{Materials Research Center for Element Strategy, Tokyo Institute of Technology, Yokohama 226-8503, Japan}

\date{\today}
\begin{abstract}
Fermi surface nesting, as a peculiar reciprocal space feature, is not only closely correlated with the real space superstructure, but also directly modulates the underlying electronic behavior. In this work, we elucidate the Fermi surface nesting feature of the IrSb compound with buckled-honeycomb-vacancy (BHV) ordering through Rh and Sn doping, and its correlation with structure and electronic state evolution. The advantageous substitution of atom sites (i.e., Rh on the Ir sites, Sn on the Sb sites, respectively), rather than the direct occupation of vacancies, induces the collapse of BHV order and the emergence of superconductivity. The distinct superconducting behavior of Rh and Sn incorporated systems are ascribed to the mismatch of Fermi surface nesting in the Sn case. 
\end{abstract}
\maketitle

\section{\label{sec:level1}\expandafter{\romannumeral1}. introduction}
 Fermi surface nesting describes the connection between different segments in Brillouin zone via a certain reciprocal lattice vector, where electrons could pick up a “free” momentum transfer equal to the nesting vector and make the energy of the system collectively unfavorable. As a result, charge (CDW) or spin (SDW) density waves emerge from effective period doubling of the crystal structure and generally act as a response of the Fermi liquid to avert the nesting instability\cite{1}. Several other alternative mechanisms are also possible to make the fermion interaction marginal by opening up an energy gap, such as quantum hall states, Mott-ness related transitions, and Cooper pairs. Tipping the delicate balance of the Fermi surface nesting, switching among several ground states can be anticipated, for example, the emergence of superconductivity from SDW order in iron pnictides\cite{2,3} or the CDW order in NbSe$_2$\cite{4} and IrTe$_2$\cite{5}. 
 
\begin{figure*}[tp]
	\includegraphics[clip,width=17cm]{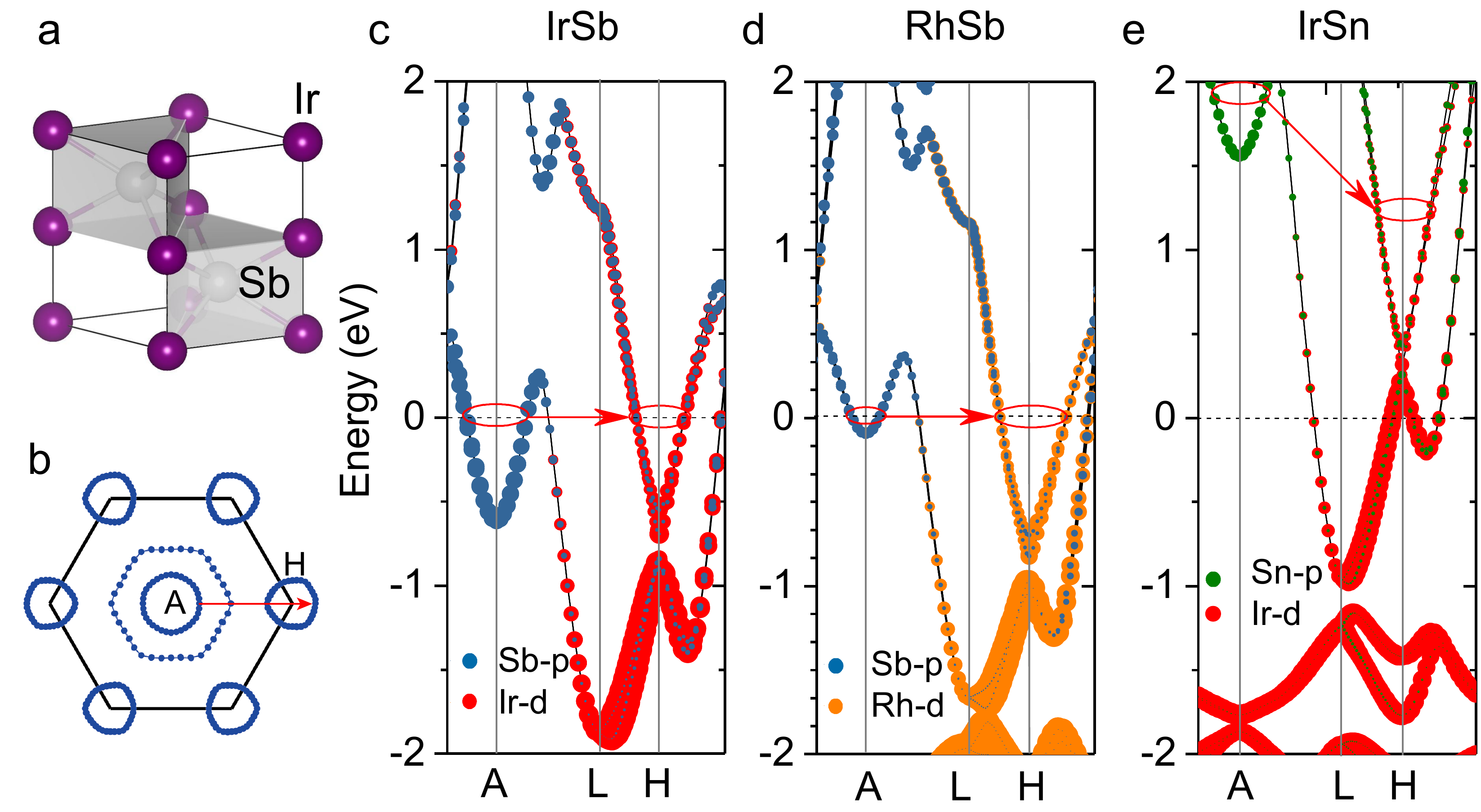}
	\caption{\label{fig1}  (a) Crystal structure of IrSb primary cell. (b) The 2D projection of the Fermi surface topology in the k$\rm_Z$ = $\pi$ plane. (c-e) Orbital-resolved band structure of NiAs-type IrSb, RhSb, and IrSn along the A-H direction. The red arrow highlights the nesting direction connecting two electron pockets around $A$ and $H$ points. }
\end{figure*}

Very recently, we reported a unique superstructural pattern of Ir$_{16}$Sb$_{18}$ consisting of 3×3×1 IrSb building blocks (Fig. \ref{fig1}a) with buckled-honeycomb-vacancy (BHV) ordering\cite{6}. In reciprocal space, this multifold enlargement coincides well with the Fermi surface nesting of $A$ to $H$ point (Fig. \ref{fig1}b and c) through a nesting vector k$\rm_N$= (1/3, 1/3, 0) and the divergence in the real part of electron susceptibility at k$\rm_N$. Compared with IrSb primary cell and other vacancy configurations, BHV ordering could effectively depress the density of states (DOS) at the Fermi level (E$\rm_F$) and hamper the transport of charge carriers. Indeed, inserting extra Ir atoms to the vacancy sites by the high-pressure method successfully suppresses the BHV ordering and induces superconductivity. A continuous $T_{\mathrm{c}}$ dome can be also realized by the incorporation of Rh atoms which is isovalent to Ir. The strong correlation of vanishment of BHV ordering and emergence of superconductivity are well-established, thus we have obtained clear evidence that the BHV ordering system is a parent phase for superconductivity. However, the explicit role of Fermi surface nesting towards the formation of BHV ordering and the emergence of superconductivity is not clear so far. It is thus imperative to scrutinize the exact vacancy-ordering-broken route and the arising of superconductivity. 

Revisiting the band structure of IrSb (Fig. \ref{fig1}c), We noticed that the bands at $A$ and $H$ points are dominated by Sb’s $4p$ orbital and Ir’s $5d$ orbital, respectively. This observation provides us an ideal opportunity to tune each electron pocket, and thus the nesting degree and its correlated phenomena. Rather than isoelectronic doping of Rh, manipulation of the component of $p$ orbital may create such a mismatch. Moreover, the introduction of extra charge carriers, as an indispensable tuning knob of superconductivity, has not been exploited yet. In this paper, we successfully introduced Sn dopant into the IrSb compounds and compared the physical properties with its counterpart of Rh doping. Distinct effects on both vacancy ordering and transport properties were revealed. The present results put the Fermi surface nesting at the root for the emergence of both superconductivity and the peculiar BHV ordering. The mechanism of $p$-$d$ orbitals nesting may be quite different from that of the conventional BCS theory, as well as other unconventional ones in cuprates and iron pnictides.

\section{\label{sec:level1}\expandafter{\romannumeral2}. experiment}
 Density functional theory (DFT) calculations were carried out by the projector augmented wave (PAW) method encoded in Vienna Ab Initio Simulation Package (VASP)\cite{7,8,9} to describe the wave functions near the core and the generalized gradient approximation (GGA) used in the Perdew-Burke-Ernzerhof (PBE) parameterization to determine the electron exchange-correlation functional\cite{10}. The structures were relaxed with a plane wave cutoff energy of 450 eV, and the forces were minimized to less than 0.01 eV/Å. The number of k-points in the Monkhorst-Pack scheme\cite{11} was 4×4×4 for structure relaxation and 6×6×6 for self-consistent calculations. Spin-orbit coupling (SOC) has been included throughout the calculations. For the projected Fermi surface nesting, 32 maximum localized Wannier functions including Ir $d$-orbitals and Sb $p$-orbitals were used in the construction of the tight-binding model.

 High purity Ir, Sn, Rh, and Sb powder were weighted with the mole ratio of Ir$_{1-x}$Rh$_x$Sb and IrSb$_{1-x}$Sn$_x$, and thoroughly mixed in a glove box. The samples were then pressed into pellets for arc melting. Powder x-ray diffraction (XRD) patterns were obtained by using a Bruker D8 Advance diffractometer with Cu-${K\alpha}$ radiation at room temperature. Transport properties were measured by using a physical property measurement system (PPMS, Quantum Design) equipped with the He$_3$ option. High-angle annular dark-field (HAADF) imaging were carried out on JEM-ARM200F.

\begin{figure*}[tp]
	\includegraphics[clip,width=17cm]{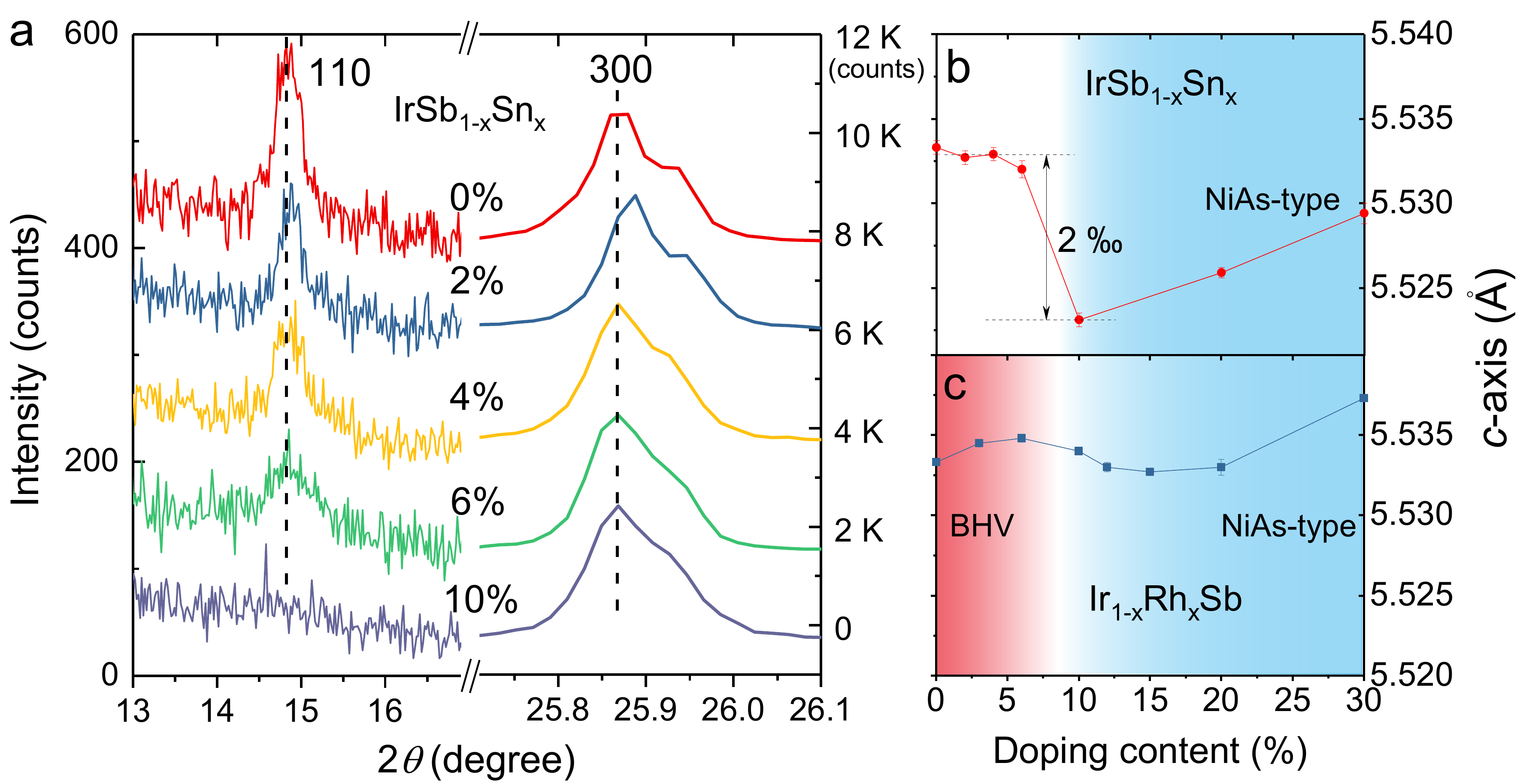}
	\caption{\label{fig2}  (a) Intensity of the Laue 110 (superstructure) and 300 (main structure) reflections at various Sn contents. The curves are vertically shifted for clarity. (b-c) Evolution of the $c$-axis lattice parameters for Ir$_{1-x}$Rh$_x$Sb and IrSb$_{1-x}$Sn$_x$ from x = 0$\%$ to 30$\%$. }
\end{figure*}

\section{\label{sec:level1}\expandafter{\romannumeral3}. results}
 It is informative to firstly check the extreme cases of Rh and Sn incoporation, namely RhSb and IrSn. Figure \ref{fig1}(d) and \ref{fig1}(e) show the band structures along the high-symmetric path for RhSb and IrSn, respectively. Although RhSb prefers to adopt the MnP-type structure, we still impose an IrSb-type primary cell for comparison. As shown in \ref{fig1}(d), the bands of complete substitution of Rh for Ir still shows rough conservation of the nesting vector, where the electron pocket at $H$ point sinks a little and others have been slightly shifted upwards. This minor change of the band structures may originate from the lattice contraction in RhSb. With one electron less in the $p$ orbital compared with Sb, the Sn incorporation introduces holes to the IrSb compound. In contrast with the band structure of RhSb and pristine IrSb, it can be seen that the band elevation of IrSn near the $A$ point is more remarkable than that near the $H$ point, leading to a mismatch of the (1/3, 1/3, 0) nesting vector. It can be anticipated that at certain doping content in IrSb$_{1-x}$Sn$_x$, the electron pocket at $A$ point is completely lifted above E$\rm_F$ to generate a Lifshitz transition. Such dramatic differences of band features in Rh and Sn doping should be reflected in several aspects of their structural evolution and physical properties. 

\begin{figure*}[tp]
	\includegraphics[clip,width=17cm]{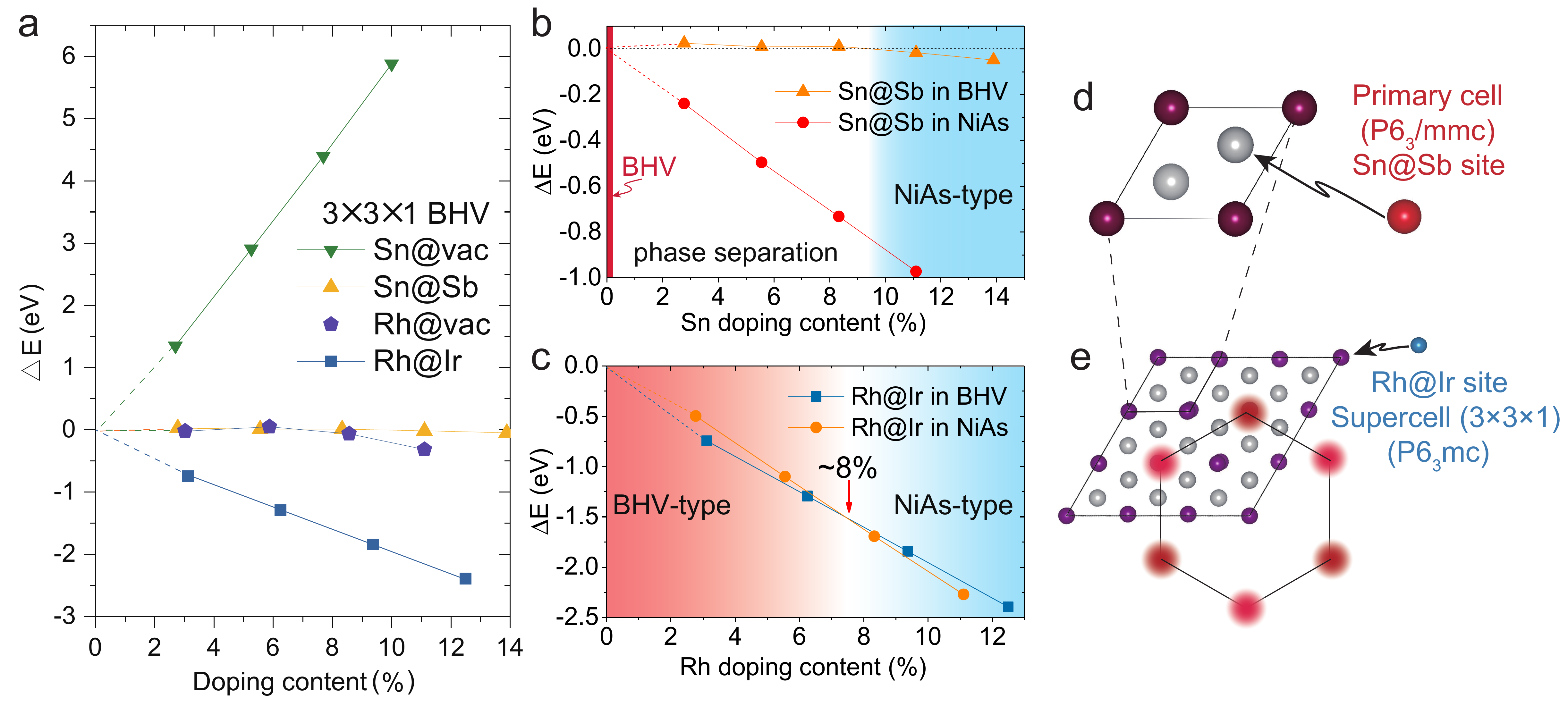}
	\caption{\label{fig3}  (a) Formation energy of Sn and Rh incorporated at different sites based on a 3×3×1 BHV superstructure. (b, c) Formation energy of Sn and Rh incorporated at anion and cation sites of a fully occupied NiAs-type primary cell, together with part of the data from (a). The calculated values based on the NiAs-type primary cell have been multiplied by 9 (3×3×1) for comparation. Red and blue shadow areas denote the region of BHV- or NiAs-type monophase. (d, e) Destruction of the vacancy ordering by the incorporation of Rh (d) and Sn (e) atoms.}
\end{figure*}

 Motivated by the above calculation, we synthesized a series of IrSb$_{1-x}$Sn$_x$ (x = 0, 0.02, 0.04, 0.06, 0.1, 0.2, 0.3) with their X-ray diffraction patterns included in Fig. S1. The low angle section (Fig. \ref{fig2}a) clearly presents the gradual suppression of the superstructural peaks of 110, while the 300 peaks from the main structure remain intact. This trend, at a first glance, resembles that of the Rh-doped case. To compare the detailed structural evolutions of the Sn and Rh-doping cases, we further extract the lattice parameters along the $c$- axis and summarize the values in Fig. \ref{fig1}b and \ref{fig1}c (evolution of lattice parameters along the $a$-axis can be found in Fig. S2). In contrast to the smooth variation in Rh cases, an abrupt drop of the $c$-axis by 2 ‰ can be observed in the Sn doped systems, indicating the breaking of the BHV order. The collapse of the $c$-axis is also observed in the previous study when the Ir$_{16}$Sb$_{18}$ compound is synthesized under pressure, but the mechanism has not been well understood due to the difficulty in tuning the synthesis pressure (especially in the low-pressure region). Considering the fine controllability of Sn in nominal composition, the abrupt drop observed in Sn-doping (and the similar trend in high-pressure samples, ref. 6) should be intrinsic.
 
 We try to elucidate this anomalous lattice collapse from the occupation of the dopants perspective and tackle this problem by considering the formation energy of Rh and Sn occupying the four possible configurations, i.e., Ir , Sb, vacancy, and interstitial sites. The $\Delta$E is defined as the free energy change accompanying Rh or Sn incorporation, wherein a negative value indicates an favorable occupancy. The results are summarized in Fig. \ref{fig3}a. Since the formation energies of anti-site occupation (i.e., Sn@Ir and Rh@Sb) and the interstitial occupation are quite high, they are not listed here for comparison. Theoretical calculations indicate that Rh atom prefers to occupy the cation site (Ir) rather than the vacancy site, agreeing well with our previous speculation. Whereas the incorporated Sn does not show an apparent energetic advantage at any site of the superstructure. The energy gain from Sb/Sn alloying is negligible in the framework of the BHV superstructure. This drives us to further check a different Sn-doping strategy by using the NiAs-type primary cell as a matrix. Figure \ref{fig3}b compares the anion substitution of Sn in both BHV superstructure and NiAs-type primary cell. As illustrated, a monotonic energy decrease can be observed for the substitution of Sn at the Sb site in the NiAs-type primary cell. The apparent negative formation energy even at 3.1$\%$ of Sn doping (Ir$_{32}$Sb$_{31}$Sn and the trend) implies that the introduction of Sn leads to phase separation of IrSb$_{1-x}$Sn$_x$ and Ir$_{16}$Sb$_{18}$ phases at the lowest doping content.
 
\begin{figure}[tp]
	\includegraphics[clip,width=8.8cm]{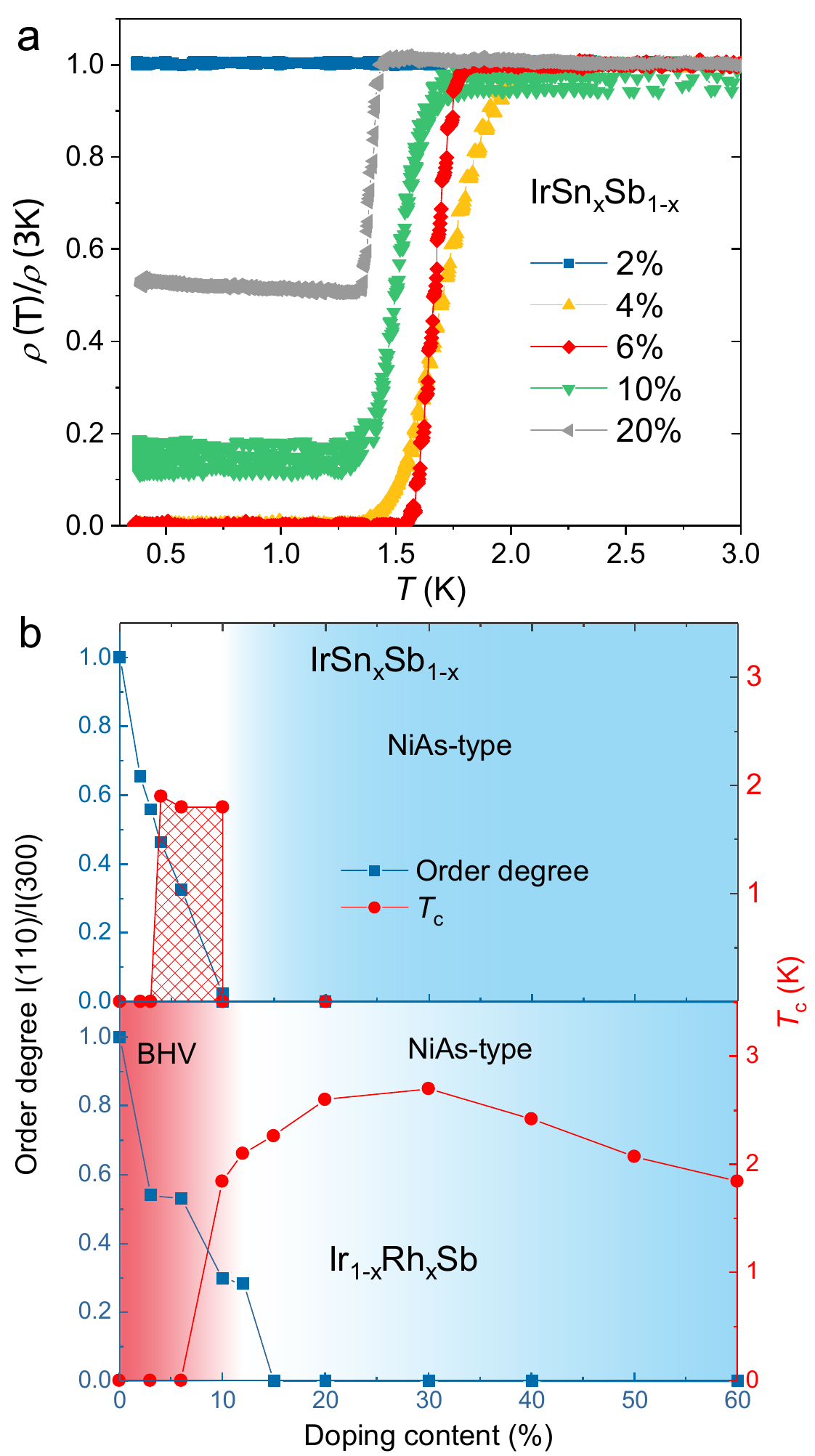}
	\caption{\label{fig4}  (a) Temperature-dependent resistivity of IrSb$_{1-x}$Sn$_x$ from 0.3K to 3K. (b) Phase diagram of IrSb$_{1-x}$Sn$_x$ and Ir$_{1-x}$Rh$_x$Sb. The order degree is defined as the normalized peak ratio of 110/300. The $T_{\mathrm{c}}$ evolution of Ir$_{1-x}$Rh$_x$Sb is extracted from Ref. 6. We note that IrSb$_{0.9}$Sn$_{0.1}$ cannot reach zero resistance, indicating a filamentary superconductivity with x \textgreater 10\%.}
\end{figure}

 The cation substitution (Rh) within a NiAs-type primary cell is also conducted and summarized in Fig. \ref{fig3}c. A crossover at around 8\% can be distinguished, below which the BHV superstructure is energetically more favorable. With the increase of the doping content, the BHV structure goes through a continuous phase transition to the NiAs-like primary cell, corresponding to the observed smooth variation of lattice parameters in Fig. \ref{fig2}c. On the contrary, a small portion of Sn is sufficient to destroy the BHV structure, leading to a phase separation with coexisting NiAs-type primary cell upon doping. This observation well explains the abrupt drop in the $c$-axis. Equipped with the above calculations, we could understand the distinct behavior of lattice parameters of Rh and Sn doping shown in Fig. \ref{fig2}. The abrupt drop observed for the high-pressure synthesized Ir$_{1-x}$Sb samples\cite{6} should have a similar origin as the Sn doping case. 

 The specific atomic locations of Sn and Rh are clarified and summarized in Fig. \ref{fig3}d and \ref{fig3}e, respectively. Namely, Rh cation tends to dope into the BHV supercell and replace the Ir atom, while Sn anion will substitute at the Sb site of the framework of IrSb primary cell, contrasting to our previous result of filling the vacancies by extra Ir under high pressure\cite{6}. Considering the prerequisite of the calculation of Ir$_{1-x}$Rh$_x$Sb based on a 3×3×1 superstructure, it should be a synergistic effect of Fermi surface nesting and vacancy formation energy for the observation of BHV ordering. Once such nesting is absent, the structure experiences a first-order transition from P6$_3$mc to P6$_3$/mmc as evidenced in IrSb$_{1-x}$Sn$_x$.

 Since Sn doping could effectively suppress the BHV ordering, it is natural to check its influence on superconductivity. The temperature-dependent resistivity down to 0.3 K is shown in Fig. \ref{fig4}a and their $T_{\mathrm{c}}$s are extracted in Fig. \ref{fig4}b. We can see that IrSb$_{0.9}$Sn$_{0.1}$ with apparent residual resistance sits at the boundary of bulk superconductivity, agreeing well with the disappearance of the superstructure peaks and the lattice collapse shown in Fig. \ref{fig2}. The phase diagram of Ir$_{1-x}$Rh$_x$Sb has been included for comparison. The phase diagram of the Sn-doping case is quite different from that of Rh doping. Instead of a continuous $T_{\mathrm{c}}$ evolution in Ir$_{1-x}$Rh$_x$Sb (a broad ‘dome’ structure), the $T_{\mathrm{c}}$ of IrSb$_{0.9}$Sn$_{0.1}$ remains almost constant at around 1.5 $\sim$ 1.6 K. As demonstrated in Fig. \ref{fig4}b, the bulk superconducting region is quite small (less than 10\%). These observations are reminiscent of the K$_x$Fe$_{2-x}$Se$_2$ where the $T_{\mathrm{c}}$ and the composition are barely tunable\cite{12,13,14,15,16,17,18}.
 
 Our previous work suggests the nesting vector of (1/3, 1/3, 0) to be responsible for the emergence of superconductivity without providing decisive evidence. Here, the Sn doping case comes as a necessary complement. With one less electron by the substitution of Sn, the $p$ band at $A$ point (Fig. \ref{fig1}e) quickly shift upwards, leading the nesting picture invalid. Concomitantly, only marginal superconductivity can be observed in the vicinity of the phase boundary of the BHV and primary IrSb phases. The nesting vector k$\rm_N$ plays an important role in the superconductivity in IrSb. On the other hand, with the robust Fermi surface nesting in Ir$_{1-x}$Rh$_x$Sb, the direct competition between BHV ordering and superconductivity manifests itself in ref 6. It is worth pointing that the electron pair nesting from p and d bands of distinct element species (shown in Fig. \ref{fig1}c) is quite rare, as the on-site nesting from different orbits of the same elements are more common in cuprate, iron-based, and other conventional BCS-type superconductors. The non-local interaction of the electrons in Ir and Sb/Sn bands is worth further investigation to elucidate the detail of the superconducting pairing mechanism.

\section{\label{sec:level1}\expandafter{\romannumeral4}. Conclusion}
 In this work, we uncovered a distinct doping effect of Rh and Sn on the structure and superconductivity evolutions of the IrSb compound. Theoretical simulation demonstrates that the collapse of BHV order in both Rh and Sn doping cases are owing to the substitution of atom sites (i.e., Rh on the Ir sites, Sn on the Sb sites, respectively), rather than the direct occupation of vacancies. Though Sn doping also introduces superconductivity, the $T_{\mathrm{c}}$ is almost constant at around 1.6 K in a narrow range of doping concentration, and the superconductivity was quickly destroyed with the increase of the dopant, which is in contrast to the broad dome straddling across large doping content in the Rh case. The different manifestations of Rh and Sn doping can be ascribed to their distinct influences on the Fermi surface nesting. Further incorporation of magnetic elements into the vacancy site while monitoring their influence toward the Fermi surface nesting and superconductivity may reveal unforeseeable physics. This exploration is promising to enrich the arsenal of defect engineering to modify other superstructures and induce emerging phenomena. 
\section{\label{sec:level1}Acknowledgments}
This work is supported by the MoSTStrategic International Cooperation in Science, Technology and Innovation Key Program (No. 2018YFE0202600), the National Natural Science Foundation of China (51922105, 51772322, U1932217, 11974246 and 12004252), National Key R\&D Program of China (Grant No. 2018YFA0704300), the Natural Science Foundation of Shanghai (Grant No. 19ZR1477300) and the Science and Technology Commission of Shanghai Municipality (19JC1413900). The research at TIT was supported by MEXT Elements Strategy Initiative to form Core Research Center. (No. JPMXP0112101001). \\

\noindent
$^\dagger$ These authors contribute equally\\
Y.Q.: qiyp@shanghaitech.edu.cn\\
X.C.: chenx29@iphy.ac.cn\\
H.H.: hosono@mces.titech.ac.jp\\

\end{document}